\documentclass[preprintnumbers,amsmath,amssymb,showkeys]{revtex4}
\usepackage{diagrams}
\input{diagrams}
\usepackage{graphicx}
\usepackage{dcolumn}
\usepackage{bm}

\newtheorem{lemma}{Lemma}
\newtheorem{corollary}{Corollary}

\begin{document}

\title{Model Theory and the $AdS/CFT$ correspondence.}

\author{Jerzy Kr\'ol\footnote{Talk presented at {\it IPM String School \& Workshop}, held in Queshm Island, Iran, 05 - 14. 01.2005.}} 
\affiliation{
Institute of Physics, University of Silesia, Uniwersytecka 4, 40-007 Katowice, Poland, \\ {\tt iriking@wp.pl}\\
}%


\begin{abstract}
We give arguments that exotic smooth structures on compact and noncompact 4-manifolds are essential for some approaches to quantum gravity. We rely on the recently developed model-theoretic approach to exotic smoothness in dimension four. It is possible to conjecture that exotic $R^4$'s play fundamental role in quantum gravity similarily as standard local 4-spacetime patches do for classical general relativity. Renormalization in gravity--field theory limit of AdS/CFT correspondence is reformulated in terms of exotic $R^4$'s. We show how doubly special relativity program can be related to some model-theoretic self-dual $R^4$'s. The relevance of the structures for the Maldacena conjecture is discussed, though explicit calculations refer to the would be noncompact smooth 4-invariants based on the intuitionistic logic.  
\end{abstract}

\maketitle

\section{\label{sec-3}Introduction and Motivation}
The $AdS/CFT$ correspondence has been the subject to many substantial checks since its discovery by Maldacena \cite{maldacena1997}. The best known example of the correspondence, namely superstring on $AdS_5\times S^5$ background versus superconformal Yang Mills $SU(N)$ on 4-dimensional Minkowski space (with 4 supersymmetries), is especially well known and explicitly described \cite{witten1998}. 

The correspondence expresses the fact that two completely different theories, the theory of (quantum) gravity - sustring in 10 dimensions and superconformal pure YM theory on 4-dimensional flat space, hence a theory without gravity, are completly equivalent to physical predictions they provide. Moreover, the correspondence correlates the weak - strong (perturbative - nonperturbative) coupling limits of the theories. 

On the YM side of the correspondence a lot of effort went into obtaining less supersymmetric and nonconformal YM theory which would be as much realistic as possible. The goal would be the 4-dimensional QCD-like theory (or the Standard Model of particle physics) as completely dual to the sustring theory \cite{Cvetic2005}. This fascinating aim, however, has not been achieved yet. There exist many proposals how to proceed toward it. The purpose of this paper is to present arguments that some new mathematical tools can be relevant for such purposes. The tools in question are exotic smooth differential structures on the topologically trivial $\mathbb{R}^4$. However, one should refer to the formal mathematical objects in perspective established by the model-theoretic paradigm rather than ascribe to the absolute classical approach where various mathematical tools are placed in the absolute ''Newton-like classical'' space, and governed by the ever present absolute classical logic.  

We attempt to explain briefly reasons that model theory become relevant for the Maldacena duality or quantum gravity in general. First, the Maldacena duality, stating the equivalence of two different theories in different dimensions, is perfectly suited for applying the tools of category theory. Categories can carry their internal logic. Possibly, models of formal theories, when taken in general categories, can guide us how to describe the duality. 

Next, the attempts to build a theory of QG are faced with extremly high energies, several orders higher than in Standard Model (SM) of elementary particles or in classical general relativity (GR). The proposed theories describing these limits, the theories of QG, rely heavily on the mathematical apparatus rather than experimental data. In such a case the theories can be especially sensitive to the degree of formalization of mathematics involved. If we had experimental data concerning the QG regime, it would be enough to propose the mathematical models explaining the results. In the opposite situation, which we face with, the mathematics with its delicate and sometimes subtle problems can be decissive for the successful formulation of QG. All the more that QG unified with other interactions is sometimes referred to as the ''theory of everything'', in the sense that it should deal with any physically valid situation. What we propose is to take into account also the model theoretic aspects of formal mathematical tools involved in the formulation of QG. This is connected with the necessarily increasing degree of formalization of the theory, and hence more and more formal approach requires referring to the nonuniqueness of formal models of the theories and languages used. Thus, not only objects derived from mathematics but also formal languages and their models can become valid objects for the full theory of QG (see also \cite{benioff1999,benioff2002,benioff2003}).   

Finally, the mathematical tools involved in QG should be applicable in all the scales of energy, from big bang, black holes to the everyday's scales of energy. The mathematical formalism seems to be extremely invariant fixing the absolute point of view of the ''observer'' with his absolute mathematical tools, like classical logic and real or natural numbers, that have to maintain relevance in all those extremly different physical regimes. It would be quite a surprise if there were no need to modify some tools when trying to describe such extremly different limits properly. All the more that mathematics itself can menage the objects properly if modified at very fundamental level, provided that the theories are sufficiently formal. The suitable tools derive from model theory.

Thus, our aims are twofold. First, we try to collect arguments that indeed the absoluteness of classical logic organizing the theory at the metalevel must be substantially weakened. The logic will then shift to intuitionism. Similarly, natural and real numbers can be considered as varying objects important for the description of QG. Both may be properly grasped by referring to some tools from model theory.
Second, in the limit of varying fundamental objects as above, one can still refer to a smooth structure supporting the changes. This is exotic smooth differential structure on topological $\mathbb{R}^4$ and such a structure can be relevant for the generelized Maldacena duality. 

Let us begin with some examination of the ''axiom'' stating that the physical theory, if formulated mathematically, should be necessarily organized by classical logic at its metalevel, i.e. at the user level of this theory. 

First, the coordinate spacetime frames that the theory referes to at some of its limits are usually understood physically as determined by some measuring devices like physical rods and watches. These, in turn, when described mathematically are usually understood as segments of real numbers. It is straightforward that the real numbers associated with the rods and watches have all the properties of the formal real numbers described by some sets of formal axioms. It is also known that the real axis is uniquely described in the second order predicate language. Thus, the rods considered formally require the reference to the second order properties. This means that some rich part of Zermelo-Fraenkel set theory with the axiom of choice (ZFC) is assumed to be valid. The question is: does the use of our physical measuring objects really justify it, and if not, what alternative do we have, if such alternative exists in mathematics at all? 

Let us assume at the beginning that we have rods as physical objects but we refer directly to the points on the idealized 1-dimensional rod and we do not want to refer to the subsets of the points of this rod in the formal sense of ZFC. This seems to be  physically natural assumption. Thus, our formal language, allowing for the above presentation of the rod, should be the first order predicate language with the names-symbols for every point of the one-dimensional ideal rod. In this way we have the correspondence between points of the rods and real numbers in some segment of reals. But we do not extend it over to the corespondence between subsets in the sense that we avoid stating the properties which require the higher order language. 
    
Now, according to classical model theory, any theory in the first order predicate language has infinitely many different models of any cardinality greater than or equal to the cardinality of the language. Thus in order to retrieve uniqueness we can follow two possibilities: one is to raise the order of formal languages. But this is exactly what we have attempted to avoid. The second is to built a theory which would be invariant in a proper sense with respect to some class of the first order models. The latter is more favorable since we can maintain more physical and direct picture of measurements using rods -- watches technology, and we can follow the Einsteinian idea of general relativity as if extended over nonstandard models of flat open regions of 4-dimensional spacetime.      
 
Besides, the analogy between local frames and Boolean-valued models of ZFC in the context of QM was explicitly stated already in the early papers by Davis \cite{davis} and Takeuti \cite{takeuti1978}. Then, the ideas where put forward in the context of arbitrary toposes with natural numbers object by Bell \cite{bell1986}.
 
Now, let us take the first order models of the theory of real numbers, i.e. all the first order properties of reals should hold in the models. Let us generate the models by the ultrafilter construction modulo some nonprincipal ultrafilter. It is known that these models are elementarilly equivalent to the standard models of reals \cite{KeislerChang}. This means that all first order properties of real numbers are the same for both, standard and nonstandard models. One needs the higher order properties of real numbers to distinguish the models. However, without referring to the higher order languages, one can still attempt to formulate a physical theory which would be invariant with respect to the choice of the models of reals.  

To distinguish between the models one has to refer to the classical and formal metalevel, which is ever present in an unchanged and absolute way. This is the point which requires relativization similar to the one performed in special or general relativity regarding the absolute {\it ether} or space and time. 

However, this relativization is essentially involved in the foundations of mathematics and, as we will see, cannot be considered as merely physically valid concept. 

Let us attempt to recover the information about the above relativization which is present already in the classical approach to manifolds and their smooth structures, and the relevance of both to background dependent or independent physical theories. The theories in question are sustring theory, or GR and loop QG respectively.  

In a background dependent theory one referes to the structure of the underlying manifold (metric, local patches); in the case of sustring this is the structure of 10 (11) dimensional manifold with the fixed Minkowski metric. There are well-defined points of the manifold which one can refer to. The manifold's structure derives, in fact, from the set-like structure. Besides, the metric allows one to measure distances between the points. 

In a background independent theory we have metric as a dynamical variable and the theory should be diffeomorphism invariant with respect to the 4-dimensional diffeomorphisms. The point-like structure is not important any longer in the sense that the theory should produce quantities completely independent on the points of the manifold or distances between these. However, the quantities are sensitive to the smooth structure of the underlying manifold. Thus the description of the smooth invariants, hence the smooth structure, should not refer to the point-like distances. 

The natural question can be asked: are there smooth structures defined such that the reference to the point-like level of the underlying manifold is limited in principle, with the exception of general knowledge of the topological type of the manifold? 

Let us observe that: if we allow for the changes of the models of SET (ZFC), as defining the reals building the manifold, the changes can cause the limiting access to the point-like structure of the manifold due to the smearing of the reals between the models. 

However, in principle, we should always recover the unique point-like description by the choice of sufficiently high classical metalevel where the models can be recognized and distinguished. Thus the change of the models of ZFC affects merely the formal presentation of the set-like point structure agreeing with its topological and smooth structures. They vary altogether along with the changes of the models of ZFC. 

However, what if the change in the models of ZFC occurs but a smooth structure is fixed and, moreover, the changes of the models are essential for the description of the smooth structure?

In that case, indeed, the smooth structure cannot refer uniquely to the point-like SET structure, and this unrespecting of the level of points is essential for the description of the smooth structure. However, it would be enough to raise the metalevel to the ambient classical one such that one can distinguish the models and hence the shift in reals defining the manifold can be detected as model-dependent. Thus, the corresponding smooth structure would not be defined on a manifold with $\mathbb{R}^n$ local patches but rather the structure varies again from one model to the other. 

However, raising the metalevel cannot be controlled formally in a full degree. Thus, it is not obvious that the raising the metalevel and attempting to keep it sufficiently formal give the classical result. Besides, in the complex theory -- metatheory there are always present some informal ingredients giving the limitations for formalization \cite{Vaananen} (see Appendices A and B). 

Thus, attempting the further formalization of the complex theory -- metatheory, the intuitionistic logic appears at the metalevel rather than the classical one \cite{krol2004}. 
Moreover, the lack of formal tools makes that the different formal models of basic mathematical structures, like reals or naturals, can happen to be indistinguishable. The models exist formally, although limitations at the metalevel make them indistinguishable. The attempt to formalize fully the metalevel results in the intuitionistic logic behind mathematical structures. The weakening of the logic gives the indistinguishability of the models.

The emergence of intuitionistic logic is evident when we start to approach arbitrary smooth real manifold literally by the symbols of some formal language. Then open subsets become important. The subsets generate topology as well. To extend this kind of description over all smooth manifolds from the category of these and smooth maps between them, one has to deal with the intuitionistic logic of the {\it smooth} topos. The topos is {\it Basel} topos which was constructed in \cite{ReyesMoerdijk}.

The symbols of the formal language respecting the open covers of the manifolds appear to be interpreted already in Basel topos. This can be understood as generic situation for the emergence of the intuitionistic logic at the metalevel when formalized. 
As the consequence two different models of reals appear which cannot be formally distinguished.

We refer the reader to the Appendices A and B for more detailed explanation of these matters.

In such a way we can be ensured about the existence of some smooth differentiable structure on some smooth manifold which does not derive from the fixed SET structure of the manifold. This structure is identified with exotic smooth $R^4$ \cite{krol2004}. The point-like level of this manifold dynamically changes between some models of ZFC, and because of that the points level does not generate the structure. This is rather the dynamics of the changes of the models of ZFC. Surprisingly, such a structure is better suited for the expressing the background independent quantities, as required by GR or QG, then the referring to the standard smooth structures, which are tightly connected with the point-like structures of the smooth manifolds. Besides, the invariants of the exotic $R^4$'s might be generated by respecting the changing models of ZFC, and these invariants can be useful in the background independent formulation of QG \cite{krol2004}.        

In the next three paragraphs we present the reasons for considering exotic smooth $R^4$'s in the context of GR, QG and the Maldacena duality. The model theoretic constructions, essential for establishing the connection, are discussed in Appendices A and B.    

\section{\label{sec11}Why exotic smooth structures in QG?}
Recently Pfeiffer \cite{pfeiffer2004} has presented some arguments that exotic smooth structures, on compact as well as on noncompact 4-manifolds, should be enormously important for quantum gravity (QG) formulated in a path integral form, i.e. as a sum over geometries on compact, smooth 4-manifolds with boundaries. The relevance of exotic smooth structures on noncompact $\mathbb{R}^4$ for QG formulated as a background independent theory was conjectured already in \cite{krol2002,krol2004}. In this and subsequent paragraphs we attempt to go further in explaining the relevance of exotic smoothness in dimension 4 for QG program. In fact, we argue that exotic smooth $R^4$'s can play a significant role for QG, which is similar to the role played by standard $\mathbb{R}^4$ in classical general relativity because of its local behaviour. Coordinate axes are considered usually as the set of real numbers. Nevertheless, such entities can be formally perceived as the models of the theory of real fields. The models are unique only in the higher orders of the formal language (in fact, the second order is sufficient). This interplay between models and formal languages gives some kind of an additional degree of freedom to a theory. 

The success in generating nontrivial combinatorial 4 manifold invariants by QG, could prove essential for the pure differential geometric task of classifying and distinguishing different nondiffeomorphic smooth structures on 4-manifolds via their PL-structures. This was partially achieved by Mackaay state sums or Crane-Yetter smooth invariants \cite{mackaay1,mackaay2,crane1997}. Donaldson and Seiberg-Witten smooth invariants are known to distinguish exotic smooth structures on compact closed 4-manifolds, but it is also known that these invariants are not refined enough to distinguish between all different nondiffeomorphic smooth structures on some 4-manifolds (see \cite{stern1999} for the case of non simply connected 4-manifolds). Thus, to find 4-invariants which would be powerful enough and combinatorial is a challenge for differential geometry. Moreover, in dimension four one can have infinite continuum many nondiffeomorphic different smooth structures on noncompact 4-manifolds. They also have their PL structures associated uniquely, so to calculate suitable invariants in this case is again challenging \cite{taylor1997}.  

Since the ''early'' days of the development of topological quantum field theories (TQFT) \cite{atiyah1988,atiyah1989,witten1988} it has been realized that the connection of physical quantum theories with some subtle mathematical questions regarding smoothness of 4-manifolds is very deep and nontrivial. Witten \cite{witten1988,witten1994} was able to show explicitly how one can obtain Donaldson polynomial invariants as correlation functions in certain twisted supersymmetric, $\cal{N}=$ 1 Yang-Mills theory on compact smooth 4-manifolds. This Yang-Mills theory possesses some features of the realistic QCD. Namely it is formulated in 4 dimensions, with an asymptotic freedom and chiral symmetry breaking, and with a dynamically generated mass gap. Donaldson polynomial invariants are smooth 4 compact, oriented manifold invariants invented by the nontrivial use of the Yang-Mills theory technics \cite{donaldson1990}. They should distinguish between different exotic smooth structures of compact closed oriented (simply connected) 4-manifolds. Witten and Seiberg proposed to calculate some other smooth 4-invariants, the Seiberg-Witten invariants, by the inherent use of physical ideas dealing with some dualities of the theories involved \cite{witten1994a}. The relation between Seiberg-Witten and the Donaldson invariants is mathematically nontrivial and is a subject of some profound conjectures \cite{donaldson1990}.    

The involvement of ideas from physics in the creation of the invariants is not accidental. It is higly essential and the relations are crucial for both fields: physics and mathematics. This is a vast field of investigations by itself and in this paper it is only touched upon. The main lesson is that there is crucial involvement of 4-smoothness in some physical field theories and appealing to the physical concepts is indispensable when pure mathematical questions, regarding 4-smoothness, appear. This should not be given up only on account of the presence of some technical complications.

In $3+1$ quantum gravity the connection is not less profound. GR needs 4-dimensional differential structure of 4-manifolds in the very intrinsic way. Moreover, formal partition functions as calculated in path integral formulation of QG (if it is formulated) have to be smooth invariants of the underlying compact, closed 4-manifolds \cite{pfeiffer2004}. Moreover, formal partition function of quantum general relativity as computed by purely combinatorial means is an invariant of PL 4-manifolds. PL-invariance is guranteed by an invariance with respect to the finite sequences of the Pachner moves \cite{pfeiffer2004}. For any smooth manifold of dimension $d\leq 6$ any smooth structure determines uniquely a combinatorial PL-structure and conversly, any PL-structure determines uniquely a smooth structure of d-manifold \cite{gompfStipshitz1999}. Thus, the partition function is automatically an invariant of the corresponding 4-dimensional smooth manifold. These PL-invariant partition functions generally take the form of state-sums and can be schematically represented as follows:
\begin{equation}\label{e1} 
Z=\sum_{\{colourings\}}\prod_{\{simplices\}}(amplitudes).
\end{equation}
Any colouring labels the faces of simplices from the triangulation of the manifold. Amplitudes are numbers-traces associated with any such labeling \cite{pfeiffer2004}. All mastery with construction of suitable invariants is the appropriate choice of the category where the sets of colourings derive from, and such that the invariance with respect to the Pachner moves holds. If so, one has PL, hence diffeomorphism invariance of the (\ref{e1}). We do not review here the existing invariants in dimensions 3 or 4, since there is a vast literature on this subject \cite{pfeiffer2004}, however the general observation can be made, namely while shifting from the lower to the higher dimension suitable higher n-categories have to be involved. This became clear especially after the publications of Baez and others in the context of TQFT, where a canonical functor was considered, the one from the category of $n$-cobordisms to the $n$-category of $n$-vector spaces \cite{baez1995,baez2004}. Crane and Yetter proposed to build 4-manifold invariants based on the category of finite dimensional representations of the quantum group $U_q(sl(2))$ where $q$ is a principal $4r$-th root of unity. Then, it became evident that it is possible to construct 4-manifold invariants out of some special 2-categories. Indeed, a detailed construction was performed by Mackaay \cite{mackaay1}. The 2-category he referred to was some kind of spherical 2-category (in the simplest variant it is a semi-strict 2-category of 2-Hilbert spaces which was the completely coordinatized version of 2Hilb category as introduced by Baez \cite{baez1995}). The highly categorical context, when 4-dimensional smoothness is considered, is also apparent in the model-theoretic approach to the subject. In the last section We will make some comments on the possible way of generating 4-smooth invariants via model-theoretic means.
 
Following Pfeiffer \cite{pfeiffer2004}, from the point of view of physics and QG, the existence of state-sum invariant of PL 4-manifolds would be of special interest. The invariant in the degenerate limit should yield the path integral quantization of $d=3+1$ pure general relativity. 
The degenerate limit is understood by analogy with the limit of $2+1$ Turaev-Viro state sum invariant which yields the degenerate Ponzano-Regge model. The latter deals with the background independent quantization of pure GR in 3-dimensions with the vanishing cosmological constant and with Riemanian signature of the 3-metric. Degeneracy of the limit means that the limiting partition function yielded from the Turaev-Viro invariant is divergent, however, after removing the bad terms, it is still invariant with respect to the Pachner moves  \cite{pfeiffer2004}. We will understand that in the case of $3+1$ invariants the ''regularization'' of the degenerate limit can be connected with special model-theoretic properties of exotic $R^4$'s.   

Moreover, in the context of sustring theory it was shown \cite{strominger1993} that certain scattering amplitudes of fivebranes solitons and axions in the heterotic string theory are proportional to the Donaldson polynomials. Thus, 4-dimensional exotic smoothness is rooted in the formalism of string theory as well.  

\section{\label{sec12}Deformed Special Relativity program and exotic smooth $R^4$'s}
The possibility that some exotic $R^4$'s can generate QG effects \cite{krol2004} leads to the reverse possibility that some QG effects could be compensated by a suitable choice of some exotic $R^4$'s. In this section we attempt to analyze the range of the validity of this later question by, first, dealing with the special limit of QG, the one which is explored in doubly special relativity program, and then, considering more general situations of full QG.
 
Deformed or Doubly Special Relativity (DSR) program was invented \cite{amelino2001,amelino2002}\footnote{I thank professor Giovanni Amelino-Camelia for letting me know about his papers} as a very direct and attractive attempt to find an extension of Lorentz or Poincar\'e algebra such that it would respect also a fundamental role played by Planck scale of length in the quantum gravity regime \cite{glikman2003,oriti2004}. It was assumed that the length should be invariant with respect to the extended algebra as another, next to the speed of light, invariant quantity. Then, it became clear that this requirement led to the necessity of considering noncommuting coordinates in the tangent and base spaces. In fact, spacetime coordinates are generated by translation generators in the energy-momentum space \cite{glikman2003}. The generators do not commute and they can be realized as the generators of the {\it remaining part} while the decomposition of the symmetries $SO(4,1)$ of 4-dimensional de Sitter space into Lorentz group $SO(3,1)$ and the remainder is performed.

It is known that $SO(4,1)$ being symmetry group of the 4-dimensional de Sitter space is also the conformal group of the Euclidean 3-space $\mathbb{R}^3$ \cite{petersen1999}. $SO(4,1)$ is a subgroup of $SO(5,1)$ - the isometry group of 5-dimensional de Sitter space. $SO(5,1)$, in turn, is the conformal group of the 4-dimensional Euclidean space $\mathbb{R}^4$; hence, the 4 dilations are included in the later group. Besides, 4-dimensional Lorentz group is present in the decomposition of $SO(4,1)$. In other words Lorentz group is the subgroup of the 4-dimensional conformal group. In fact, n-dim Poincar\'e group has $\frac{1}{2}n(n+1)$ generators while the conformal group in n-dimensions has $\frac{1}{2}(n+1)(n+2)$ generators. The difference is $n+1$ generators which $n$ of them generate special conformal transformations and remaining one parameter corresponds to the 4-dilations. 

It was conjectured in \cite{krol2004b} that some standard (with respect to the standard smooth structure on $\mathbb{R}^4$) contractions and, possibly, rotations of the 4-ball where some exotic $R^4$ is placed give, as the result of such localization, some non commuting ''momentum''coordinates corresponding to some noncommutative space. In fact, any arbitrarily small contraction makes the exotic smooth structure being deformed such that it cannot be embedded in the ambient standard structure extending the exotic one \cite{gompfStipshitz1999}.      

We can synthesize the whole idea by assuming that some 4-dilations of some exotic 4-space give the noncommutative space, which in some special limit gives the noncommutative 4-space exactly as in the case of DSR noncommutativity i.e. generated by four-dimensional de Sitter momentum space. The limit of QG where DSR approach is valid is the ''flat space semi-classical limit of $3+1$ quantum gravity''. This means that one takes Newton constant $G\rightarrow 0$ and Planck constant $\-h\rightarrow 0$ but such that their common ratio $\sqrt{\frac{G}{h}}$ remains finite. This finite limiting ratio is in fact Planck length and the vanishing $G$ limit of GR is flat spacetime solution of topological field theory; taking $h\rightarrow 0$ is the weak coupling limit of QG. The finite fraction $\lim_{G,h\rightarrow 0}{\sqrt{\frac{G}{h}}}=\kappa^{-1}$ corresponds to $\kappa $ which is exactly the radius of the 4-de Sitter space defined in the 5-dimensional Euclidean space as a surface: $\kappa^2=-\eta_0^2 +\eta_1^2 +\eta_2^2 +\eta_3^2 +\eta_4^2$ where $\eta_0,\eta_1,\eta_2,\eta_3,\eta_4$ are coordinates in the Euklidean space \cite{glikman2003}.       

Although the relation between some contracted exotic $R^4$ and noncommutative spaces was observed to exist, the relation does not decide the exact shape of the noncommutative space. However, it is plausible that the class of noncommutative spaces associated with contracted exotic $R^4$'s appears to be wide enough to state the following general conjecture:

A1.\label{conA1} {\it In the flat space semi-classical limit of $3+1$ quantum gravity \cite{glikman2003} one reaches the noncommutative DSR space, as connected with some reference frame of the momentum 4-de Sitter space, which is exactly, possibly deformed, the noncommutative space generated by some contracted exotic $R^4$ \cite{krol2004b}.}

or 

A2.\label{conA2} {\it Any quantum gravitational effect in the flat space semi-classical limit of QG can be locally compensated by a suitable choice of some exotic smooth $R^4$.}

Let us develop some arguments that would be relevant to another stronger hypothesis, namely the one stating that exotic $R^4$'s can be involved more essentially in a formulation of full QG than only in the limit as above.
\begin{itemize}
\item First, exotic $R^4$'s are tools suitable for the detection or measurement of exotic 4-dimensional smoothness in the case of compact 4-manifolds. This is precisely the way followed by Taylor \cite{taylor1997} in generating the smooth invariants of compact or noncompact 4-manifolds. However, compact exotic smoothness is essential for QG, at least in the path integral formulation \cite{pfeiffer2004}, hence, also some exotic noncompact $R^4$'s should be considered important there. 
\item Second, spin networks or state sums are natural states of loop QG which is a background independent formulation of QG in 4 dimensions \cite{baez1995}. Besides, natural candidates for smooth PL invariants of compact 4-manifolds are state sum invariants \cite{mackaay1}. In the knowledge that exotic smooth $R^4$'s are related to the compact smoothness as in the previous point, one comprehends that exotic $R^4$'s should be important for a background independent formulation of QG.
\item Third, considering exotic world volumes of $D3$-brabnes in sustring theory, exotic $R^4$'s may have a meaning in the perturbative sustring description of QG \cite{krol2004}. This can be further exploited in the context of the Maldacena conjecture \cite{krol2004}.    
\item Fourth, exotic $R^4$'s smooth invariants, if calculated by some model theoretic means (MTSD networks), can compansate infinities or divergencies (see section \ref{sec-1+1}). This, in turn, can help to understand perturbative, though, non renormalized, quantum gravity.    
\end{itemize}

Moreover, let us examine some general arguments tackling symmetries involved in the DSR and contracted exotic $R^4$'s.
The standard contractions of exotic $R^4$'s require reference to the 4 dimensional contractions which are elements of the conformal group of 4-space $\mathbb{R}^4$. This last group is $SO(5,1)$ or in the case of Minkowski's metric the conformal group is $SO(4,2)$. In turn, $SO(4,2)$ is the isometry group of anti-de Sitter 5-space, namely $AdS_5$. Thus, to compensate properly QG effects in the flat space semi-classical limit of QG in terms of exotic self-dual smooth $R^4$'s one should deal with a broader group than $SO(4,1)$. This broader group has to contain 4-dilations of Euclidean or Minkowski space; this might be $SO(5,1)$ or $SO(4,2)$. Any field theory, respecting these groups of symmetries and being formulated on flat, say Minkowski $\mathbb{R}^4$, while the standard smoothness of the underlaying $\mathbb{R}^4$ patch is changed to the exotic one, should contain a reference to the quantum gravitational regime as well. If the field theory on standard Minkowski 4-space respects the $SO(4,2)$ symmetry, this theory is clearly conformally invariant. Now if one switches the smooth structure to the exotic one, the field theory will not be conformally invariant any longer, since in the regime of suitable contractions the noncommutative (quantum gravitational) effects appear, or in terms of exotic smooth structure the exotic metric is not in the conformal class of the Minkowski metric. Moreover, the exotic metric need not be standard smooth one as well as the Minkowski metric is not a smooth metric in the exotic structure.

Now if one considers the basic case of the Maldacena conjecture, namely the case of sustrings on $AdS_5\times S^5$ dual to the $SU(N)$ Yang-Mills superconformal with $\cal{N}=$ 4 and $N$ large, he is faced with the possibility that the conformal invariance of the theory should be broken due to the possible intervention of exotic $R^4$ structures \cite{krol2004}. Let us observe that the $AdS_5$ is just the space whose isometry group supports conformal transformations of 4-Minkowski space and this was the initial test of the duality. Moreover, the breaking of conformal invariance in this context should be done in order to have the semi-classical limit of 4 dim QG well adapted.  	 

Thus, we can see that the involvement of exotic smooth $R^4$'s is not only limited to the flat space semi-classical limit of $3+1$ QG, but it may play more profound role in full QG. The observations encourage us to formulate the following conjecture:

A3.\label{conA3} {\it Any quantum gravitational effect in 4-dimensions, perturbative or non-perturbative, can be compensated locally by a suitable choice of some exotic $R^4$.}  

The family of exotic $R^4$'s emerges, such that $R^4$'s compensate QG effects locally. This family remains in some analogy to differentiable manifolds and its coordinate local $\mathbb{R}^4$ patches, as in the case of classical gravity. However, this family does not constitute any cover of differentiable manifold in the strict, classical sense. 

We have presented only some heuristic, general arguments for the above conjecture, however, one can become more convinced about its validity by a more careful examination of model-theoretic self-dual exotic smooth structures on $\mathbb{R}^4$. We will see that the model-theoretic approach to exotic smoothness resembles to a certain degree the philosophy behind the general relativity approach to the classical gravity. Certainly, the very meaning of this conjecture requires explanation. At our disposal these are two possible scenarios of studying the exotic $R^4$'s and their meaning in physics. One is external, based on general relations to some other well-known concepts or tools and the second internal, and it attempts to grasp exoticness from the point of view of its intrinsic nonclassical perspective (see Apendices A and B). It is strange and unexpected that the latter is more suitable for exhibiting the physical meaning of the exotic structures. On the other hand an impasse in a direct, analytical approach to exotic smoothness in dimension four caused the latter to emerge \cite{krol2002,krol2004,krol2004b}.

\section{\label{sec1-1}The general features of exotic $R^4$'s indicating that they are well suited for General Relativity and Quantum Gravity.}
Before one attempts to obtain concrete calculations concerning exotic $R^4$'s and their connection with QG or GR it is worth examining at the problem from a more general perspective and to collect some general indications showing that the connection should be studied.     
\begin{itemize}
\item First, the conjecture \ref{conA3}.A3 resembles much the Einsteinian equivalence principle (EP) in GR. Let us have a closer look at this.  

At least two versions of classical EP exist in general relativity \cite{weinberg}. EP says that {\it at every spacetime point in an arbitrary gravitational field it is always possible to choose a ''locally inertial coordinate system'' such that, within a sufficiently small region of the point, the laws of nature take the same form as in unaccelerated Cartesian coordinate system in the absense of gravitation.}

The weak EP holds when in the above statement ''the laws'' means ''the laws of motion of freely falling particles''. Strong EP holds if we understand ''the laws of nature'' as ''{\it all} laws of nature''. 

EP is connected directly to general covariance or the general feature of Lorentzian manifolds (of any manifold in fact) where it is always possible to take a locally trivial coordinate patch. A general diffeomorphisms generate gravity via the metric and Christoffel connection which in turn results in the concept of covariant derivative. The information deriving from the covariant derivative characterizes the tangent bundle of the manifold in global. 

Now let us reformulate EP as follows: In spacetime it is always possible to choose locally standard $\mathbb{R}^4$ patch in which no gravitational effects are observed, or 

A3'. {\it The standard $\mathbb{R}^4$ patch can be chosen such that this compensates locally all classical effects of gravity.}

Now the resemblance to the quantum gravitational counterpart \ref{conA3}.A3 is striking. Both statements are deeply rooted in the differential geometrical properties of four-manifolds.

\item Second, the general invariance, or coordinate independence, with respect to 4-diffeomorphisms required by GR assures us that every decomposition of spacetime into 3 space directions and time direction is a violation of the invariance. But when this happens it can be recovered just by summing over all decompositions or by showing that the result does not depend on the specific decomposition performed. However, the general invariance or the obstruction with decomposing $\mathbb{R}^4$ into $\mathbb{R}\times\mathbb{R}\times\mathbb{R}\times\mathbb{R}$ is such a fundemental feature of GR that while reconciliation with QM is considered this has to be carefully analysed. Let us note that the above decomposition agrees with the topological understanding of coordinate axes and their smooth product is understood as compatible with the topological product. Such an implicit assumption enforces one to deal with the standard smoothness, if any, of $\mathbb{R}^4$. Moreover, another implicit assumption is made. This is the order of decomposing into coordinates and recovering the smooth structure on $\mathbb{R}^4$. This order is assumed to be irrelevant and the above actions can be performed anytime and they always result in the standard smooth structure. The result does not depend on these factors and one has always control of the compatibility of the smooth and topological structures. As a result, GR, though generally invariant and coordinate independent, is built as if it were always equivalent to the coordinate dependent formulation or, as if the passage between the two were always in reach. This low level implicit assumption is however not made in the GR spirit. 

Let us point out what exotic $R^4$'s can change regarding this point. Similarily as in GR, any choice of global smooth coordinates in the accordance with $\mathbb{R}\times\mathbb{R}\times\mathbb{R}\times\mathbb{R}$ destroys the exotic smoothness. However, its analytical restoration is not available at present. The ease with restoration of general invariance in the case of GR compatible theories as mentioned above, comparing with the dificulty, or even untractability with the explicit restoration of the exotic smoothness, should be an indicator that possibly something is overlooked in the GR/QM relations. Besides, there are arguments for the fact that one should take into account 4-dimensional exotic smoothness when QG is considered \cite{pfeiffer2004}.

On the other hand, the difficulty with the recovering of the exoticness from the standard topological coordinates shows that the topological axes are very deeply and canonically entangled in the exotic $R^4$. This is the feature required by GR. A theory on exotic $R^4$ respects the fundemental GR requirement. Moreover, contracted exotic $R^4$'s can interpret canonical noncommutative relations as in QM \cite{krol2004b} and hence, heuristically, they appear to be perfect tools for unifying QM and GR. 

What is more, it follows from \cite{krol2004b} that model-theoretic self-dual exotic $R^4$'s exhibit noncommutativity when one settles fixed model-theoretic standardness and smooth standardness of coordinates. The last one is compatible with the topological $\mathbb{R}^4$ structure. That is why fixing that the topological and smooth structures agree, results in the fact that the QM-like behaviour is not any longer present. From such a perspective, it is even essential to take into account exotic $R^4$'s in order to reconcile QM and GR.    

\item Third, it was shown by Asselmeyer \cite{asselmeyer} that exotic smooth structures on some compact 4-manifolds generate distribution-like sources of gravity. Besides, Brans \cite{brans1994} conjectured that exotic smooth $R^4$ can act as sources of gravitational field. It was S{\l}adkowski \cite{sladkowski2001} showed that, indeed, some exotic $R^4$'s can generate highly nontrivial solutions of Einstein equations even in the empty but exotic $R^4$. The sources of gravity are exhibited while the exotic smooth structure is seen from the standard one, so this shift of the structures switches on the classical sources of gravity. In fact, it was proposed \cite{krol2002,krol2004} that such a shift can switch on some QG effects. This is also expressed in \ref{conA3}.A3. 

\item Forth, when the background independence issue is considered in QG, the canonical role played by exotic $R^4$'s can be noticed \cite{krol2004}. It relies on the model-theoretic approach to exotic smoothness. The point is that the background manifold of the theory is frozen unless we take the model-theoretic dimension of the objects. Owing to this new perspective the manifold can vary. This dimension respects the model extensions of standard real and natural numbers. Such a situation generates model-theoretic self-dual exotic $R^4$'s and by Brans conjecture they can act as external sources of gravitational field. Consequently, model-theoretic background independence generates exotic smooth 4-structures and this has its gravitational counterpart.    
\item Fifth, assume we have QG formulated such that it respects somehow 4-dimensional exotic smoothness (e.g. path integral formulation). The classical limit in this case, at least locally, should correspond to the relation between exotic $R^4$ and standard smooth $\mathbb{R}^4$. The relation between the manifolds, at the topological level, is trivial, namely, both manifolds are topologically identical. This $\mathbb{R}^4$, while standard smooth, is just the main ingredient of classical GR in the local behaviour.
\end{itemize}

\section{\label{sec-1+1}Exotic renormalization in AdS/CFT}
Owing to the recent growth of interest in the geometry--field theory limit of the AdS/CFT correspondence, many explicit formulas have been worked out in favour of this holographic like relation \cite{skenderis2004,deharo2001}. We attempt to connect the so-called holographic renormalization procedures with the possible interventions of exotic smooth structures on $\mathbb{R}^4$. The analysis makes an essential use of the model-theoretic self-duality of exotic $R^4$'s.

The general strategy is as follows:
\begin{itemize}
\item[1.] Localize divergent quantities, as usual, with respect to some standard smooth and model-theoretic standard (MT standard) patch $\mathbb{R}^4$.
\item[2.] Shift the patch to the one which is Robinsonnian nonstandard \cite{robinson1964} and the divergencies can be considered as nonstandard reals or nonstandard expressions in general. 
\item[3.] Change the model-theoretic environment to the intuitionistic one  where the nonstandard reals or expressions can be considered as {\it smooth}-finite ($s$-finite) \cite{ReyesMoerdijk} in the sense of the internal logic of Basel topos ${\cal B}$ (see section (\ref{sec-12})).
\item[4.] Refer to the fact that there may exist model-theoretic self-dual (MTSD) exotic $R^4$ which can interpret all the above steps \cite{krol2004b}.
\item[5.] Attempt to consider a theory which is invariant with respect to the {\it exotic diffeomorphisms} of the exotic $R^4$ rather than to the standard ones \cite{krol2004b}. 
\end{itemize}
Such a theory has been built in a kind of renormalization procedure automatically, in the sense that the shift $infinite \rightarrow  finite$ is legitimate with respect to the general symmetry of the theory. Moreover, owing to the Brans conjecture \cite{brans1994} and the connection of exotic $R^4$'s and QG, the exotic MTSD $R^4$ suggests a connection with gravity (in both classical and quantum regimes) while the above renormalization scheme is been considered. 

Since many explicit expressions heve been worked out in the holographic renormalization program we can also try to be more specific with our model-theoretic renormalization approach, at least in the geometry - field theory correspondence.
  
The conformal boundary of $AdS_5$ is $S^4$ in the Euclidean picture of the $AdS_5$ space \cite{witten1998} (This boundary can be considered as $\mathbb{R}^4\cup \{\infty\}$). Potentially, $\mathbb{R}^4$ can carry an infinite number of different smooth structures but after the one--point compactification they might be unique again (up to the smooth 4-dim Poincar\'e conjecture which is still open \cite{gompfStipshitz1999}). 

Now, let us consider $S^4$ as the one-point compactification of $\mathbb{R}^4$, but in the sense of Robinson's nonstandard analysis \cite{robinson1964}; this compactification is the quotient of the nonstandard exstension, ${\bf ^\star R^4}$, by some equivalence relation (\cite{nan1985}, p. 159).

For this reason divergent functions on $S^4$ might be considered as nonstandard functions on ${\bf ^\star R^4}$ up to the equivalence relation defining the one point compactification. 

Now, let us consider the gravity counterterms to the $AdS_5$ Einstein-Hilbert (EH) supergravity action, which have been calculated and presented in several papers (see e.g. \cite{skenderis2001,skenderis2002,skenderis2004}). These counterterms are needed to compensate for the divergencies in the EH action. The action is formulated in the bulk which in this case is $AdS_5$. 
EH action supplemented by a surface term, on anti-de-Sitter $n$-dimensional space, can be written as 

\begin{equation}\label{e2}
S_{\rm{bulk}}+S_{\rm{surf}}=-{\frac{1}{16\pi G}}\int_{AdS}d^{n+1}x\sqrt{\tilde{g}} \left(R+\frac{n(n-1)}{l^2}\right)-\frac{1}{8\pi G} \int_{\partial(AdS)}d^{n}x\sqrt{h}K
\end{equation}
The bulk term is just EH action with the cosmological constant $\Lambda = -\frac{n(n-1)}{2l^2}$, $l$ is the radius of the $AdS$ space, $K$ is the trace of the extrinsic curvature while embedding of the boundary $\partial(AdS)$ in the $AdS$ is being considered. The boundary term guarantees that when field equations are calculated with respect to the above action (i.e. the variation with respect to the metric fixed at the boundary via its normal derivatives is calculated) they become Einstein equations \cite{emparan1999}. $h$ is the metric induced at the boundary from the bulk metric $g$ in the above sense. The action (\ref{e2}) is divergent, since the bulk is noncompact $AdS$ space and hence it has infinite volume, and the boundary term is divergent since the induced metric $h$ diverges on the boundary. Nevertheless, one can extract precise divergent counterterms. They are functionals of the boundary curvature and its derivatives only, and the counterterms depend on the radius $l$. 

In the specific limit of the AdS/CFT correspondence we are interested in in the context of holographic renormalization, the classical supergravity theory in the asymptotic bulk $AAdS_5\times S^5$ is equivalent to the quantum field theory on the conformal 4-boundary. $AAdS$ is asymptotically anti-de Sitter 5-space \cite{skenderis2002}. Having the radial bulk coordinate $\rho$ such that $\rho=0$ corresponds to the asymptotic boundary, we can regularize the action taking surface corresponding to $\rho=\epsilon$ \cite{skenderis2001a}. Thus, the regularized action can be written as \cite{skenderis2002}
\begin{equation}\label{e3}
S_{\rm{reg}}={\frac{1}{16\pi G}}\int_{\rho\geq 0}d^{4+1}x\sqrt{\tilde{g}} \left(R+\frac{6}{l^2}\right)-\frac{1}{8\pi G} \int_{\rho=\epsilon}d^{4}x\sqrt{h}K
\end{equation}
Counterterms of the action in the case of $AdS_n$ space are 
\begin{eqnarray}\label{e4}
&S_{\rm{ct}}=\frac{1}{8\pi G}\int_{\partial(AdS)}d^{n}x\sqrt{h}F(l,{\cal{R}},\nabla{\cal{R}})=\\\nonumber
&\frac{1}{8\pi G}\int_{\partial(AdS)}d^{n}x\sqrt{h}\left[\frac{3}{l}+\frac{l}{2(n-2)}{\cal{R}}+\frac{l^3}{2(n-4)(n-2)^2}\left({\cal{R}}_{ab}{\cal{R}}^{ab}-\frac{n}{4(n-1)}{\cal{R}}^2\right)+...\right].
\end{eqnarray}
The counterterms corresponding to the regularized action on the $\rho=\epsilon$ surface can be written as
\begin{eqnarray}\label{e5}
S_{\rm{ct}}^{\epsilon}[g_0]= 
\frac{1}{16\pi G}\int_{\rho=\epsilon}d^{n}x\sqrt{h}\left[2(1-n)+\frac{1}{n-2}{\cal{R}}-\frac{1}{(n-4)(n-2)^2}\left({\cal{R}}_{ab}{\cal{R}}^{ab}-\frac{n}{4(n-1)}{\cal{R}}^2\right)-\log\epsilon a_{(n)}\right]\nonumber
\end{eqnarray}
where $\log\epsilon a_{(n)}$ is the logarithmic term while the $S_{\rm{reg}}$ action is expanded in powers of $\epsilon$ as \cite{skenderis2001a}
\begin{eqnarray}\label{e6}
S_{\rm{reg}}^{\epsilon} = 
\frac{1}{16\pi G}\int d^nx \sqrt{g_{(0)}}\left( \epsilon ^{-n/2}a_{(0)}+\epsilon^{-n/2+1}a_{(2)}+...+\epsilon^{-1}a_{(n-2)} 
 -\log \epsilon a_{(n)}\right) +{\cal{O}}(\epsilon_{(0)}) \nonumber
\end{eqnarray}
and the metric in the neighbourhood of the boundary in the bulk is \cite{skenderis2002}
\begin{equation}\label{e7}
ds^2=\tilde{g}_{\mu\nu}dx^{\mu}dx^{\nu}=\frac{1}{r^2}(dr^2+g_{ij}(x,r)dx^idx^j.
\end{equation}
Now the boundary is located at $r=0$ and in the limit of $r\rightarrow 0$ the metric has a smooth limit $g_{(0)ij}$.
Hence, it can be written
\begin{equation}\label{e8}
g_{ij}(x,r)=g_{(0)ij}+rg_{(1)ij}+r^2g_{(2)ij}+r^3g_{(3)ij}+...\,\,.
\end{equation} 
This limiting, boundary metric $g_{0}$ was used in (\ref{e5}) and $r$ is the defining function of the boundary metric, namely it holds \cite{skenderis2002}
\begin{equation}\label{e9}
g_{(0)}=r^2\tilde{g}_{|\partial AdS}
\end{equation}  
and the bulk metric $\tilde{g}$ is evaluated at the boundary of $AdS$.

Introducing new coordinates $\rho=r^2(=\epsilon)$ one can write \cite{skenderis2002}
\begin{eqnarray}\label{e10}
ds^2=\tilde{g}_{\mu\nu}dx^{\mu}dx^{\nu}=\frac{d\rho^2}{4\rho^2}+\frac{1}{\rho}g_{ij}(x,\rho)dx^idx^j\\
g(x,\rho)_{ij}= 
g_{(0)ij}+\rho^{1/2}g_{(1)ij}+\rho g_{(2)ij}+\rho^{3/2}g_{(3)ij}+...
+\rho^{n/2}g_{(n)ij}+\tilde{h}_{(n)ij}\rho^{n/2}\log\rho +...\rm{.}
\end{eqnarray} 
Now subtracting all divergent terms $-S_{\rm{ct}}^{\epsilon}[g_0]$ from $S_{\rm{reg}}^{\epsilon}$ in (\ref{e6}), while still staying on the regularizing surface $\rho=\epsilon$ and performing $\epsilon\rightarrow 0$ limit in which renormalized action $S_{\rm{ren}}[g_{(0)}]$ takes form \cite{skenderis2002}
\begin{eqnarray}\label{e11} 
&&S_{\rm{ren}}[g_{(0)}]= \\ \nonumber  
&&\lim_{\epsilon \rightarrow 0}\left(S_{\rm{reg}}-\frac{1}{16\pi G}\int_{\rho=\epsilon }\sqrt{h}\left[A \right] \right) \\ \nonumber
&&{\rm where}\,\,\, A=2(1-n)+\frac{1}{n-2}{\cal{R}}
-\frac{1}{(n-4)(n-2)^2}({\cal{R}}_{ij}{\cal{R}}^{ij}-\frac{n}{4(n-1)}{\cal{R}}^2)-\log \epsilon a_{(n)}.
\end{eqnarray} 
Thus, all finite number of divergent terms in the action were expressed in terms of the boundary induced metric and then, when the divergent terms were subtracted, the renormalized action is finite and well-defined. Now one can produce various boundary $n$-point functions via field theory calculations \cite{skenderis2002} and they should precisely characterize 5-dimensional bulk geometry of the asymptotic anti-de Sitter space. 

Now, following \cite{solodukhin1999}, let us approach the 5-dimensional noncompact bulk by the increasing sequence of compact embedded submanifolds with boundaries. Therefore, we have submanifolds $M_\rho$ with boundaries $\partial M_\rho $ parametrized by $\rho $ such that $M_\rho \rightarrow M$ for large $\rho $. The action functional (\ref{e2}) as written on the $M$ should be approached by the sequence $\{S_\rho \}$ of the corresponding functionals. This means that divergencies of the limiting functional are also approached.

Now instead of taking the limit $M_\rho \rightarrow M$ and $\partial M_\rho \rightarrow \partial M$ and renormalizing the action via counterterms which are the functionals of the curvature invariants at the boundary as discussed above, let us take the limit modulo some nonprincipal ultrafilter $U$ on the infinite set of indices $\{\rho \}\approx\omega$ \cite{nan1985}. As the result, we get two ingredients
\begin{itemize}
\item a nonstandard boundary $ ^\ast \partial M \approx (\{\partial M_{\rho_n}\}^{n\rightarrow \infty })/{U}$, which locally can be identified with the nonstandard ${\bf ^\star R^4}$,   
\item a nonstandard action containing some symbols which are nonstandard or which refer to some nonstandard real numbers.
\end{itemize}
The first ingredient can be considered as the generating nonstandard ${\bf ^\star R^4}$ which was used in the nonstandard construction of the one-point compactification of $\mathbb{R}^4$ at the beginning of this section. 
The second ingredient can be worked out when one takes in the renormalizing limit $\epsilon \rightarrow 0$ in (\ref{e11}) or (\ref{e6}) just sequences of $\epsilon $'s as corresponding to $\rho$'s. Then, on such a family of the expressions let us perform the modulo nonprincipal ultrafilter operation. Thus, the divergencies of the counterterms $S_{\rm{ct}}^{\epsilon}[g_0]$ as in (\ref{e5}) correspond to nonstandard reals, when the model-theoretic limit (i.e. modulo the nonprincipal ultrafilter) is taken and when the nonstandard boundary is generated as well. Moreover, the limiting action now can be seen as the nonstandard functional defined on the nonstandard functions which, in turn, are defined locally on ${\bf ^\star R^4}$ with the values in ${\bf ^\star R}$ or $ ^\ast \mathbb{C}$.         

The main point of this model-theoretic construction is that having the sequence of the submanifolds and their boundaries which carry canonical order with respect to the increasing parameter $\rho$, one does not respect this order any longer and makes the identification of the boundaries modulo some ultrafilter. This new limiting procedure is just the one which replaces the old, order-like, one. However, there is still a connection with the former limiting procedure. This is clearly the MTSD property of exotic $R^4$'s. Having the appropriate MTSD exotic $R^4$, one can choose the $\mathbb{R}^4$ patch where ''standard'' infinities of the action are detected or switched to the nonstandard ${\bf ^\star R^4}$ patch where the infinities are still real, though nonstandard, numbers. The shift between the patches can be compensated in principle by some intuitionistic environment modelled by the inside of the Basel topos. The smooth structure which can afford all the changes is just model-theoretic self-dual exotic $R^4$. Hence, instead simply dropping the infinities, one can create a theory which can model canonically the renormalization procedure. The symmetries should respect MTSD exotic diffeomorphisms \cite{krol2004b}. Thus the use of exotic $R^4$'s would be just respecting the philosophy behind general renormalization technology.

To be more specific, let us abbreviate the divergent counterterm in \ref{e11} by $A(g_0,a_n,\epsilon)$. Taking the limit of $\epsilon\rightarrow 0$ we can reexpress it by taking infinitesimal small $\epsilon $'s but the expression $A(g_0,a_n,\epsilon)$, contains the symbol referring to the nonstandard numbers, $\epsilon $, and can take nonstandard (infinite big) values: ${ ^\ast A(g_0,a_n, { ^\ast\epsilon})}$. To recover standard expression for $A$, one can take standard part operation, or $ ^\ast A$ can be embedded into the weaker logical environment, namely one deriving from the smooth Basel topos. In this intuitionistic logic infinite big numbers can be considered as {\it smooth} finite. This expresses the fact that in the Basel topos the true real numbers contain nonstandard numbers as well. The classical projection, which is not unique, gives standard expressions or Robinsonian nonstandard ones. Remembering that we are in dimension four, we can make use of the MTSD exotic smooth structures. The structures in question can be generated by the dynamically changing language $L({\rm SET},{\cal B})$ (see Appendix B). Clearly, MTSD smooth exotic structures on $\mathbb{R}^4$ are invariant with respect to the changes given by the dynamical language. Hence, divergencies as in the countertem \ref{e11} should be the building ingredient of the exotic structure. However, how to describe the structure in terms of global differential calculus is not known at present.  

The structures need the divergent terms as their building blocks and which are described as divergent in a specific, model--theoretic limit of the structures. Furthermore, the structures can compensate for the divergencies, hence one need not to subtract the terms on the 4--dimensional field theory side by hand. Instead, the theory should be formulated on exotic rather than standard 4-spaces. 

Now we can see that exotic MTSD smooth $R^4$'s localized in the asymptotic $AdS$ boundary can generate the counterterms which, in turn, can cause that the bulk 5-dimensional gravitational action becomes finite. In addition to this the 5--dimensional gravitational divergencies are translated into exotic structures on 4-space, which shows the specific connection of exotica in 4--dimension and 5--dimensional gravity; thus, not only the 4--dimensional sources of gravity are to be considered as gravitational impact of 4--exotica. Besides, these are 5--dimensional gravitational divergencies. The MTSD 4-structures seem to be well suited for the renormalization questions. 

However, more explicit calculations would require analytical tools which at the present time are out of our reach. This is the task for forthcoming studies which possibly will improve the situation. An important step towards grasping the Yang-Mills theories on some exotic $R^4$'s and towards description of exotic metrics on these, has been recently made by Kato \cite{kato2004}\footnote{I would like thank professor Robert Gompf for letting me know about this paper.} With the help of these metrics one could attempt to understand what is the role of exotic smooth $R^4$'s for the susy breaking pattern in the AdS/CFT correspondence. As a consequence, more realistic 4-dimensional YM theories can emerge as dual to the sustrings in 5 dimensions. More detailed studies of that case, however, have to wait untill exotic smooth invariants will be generated via model-theoretic methods. Although, some remarks concerning the modification of YM and sustrings sides of the Maldacena cortrespondence by exotic smooth $R^4$,s, one can find in \cite{krol2004}.        

\appendix
\section{\label{sec-11}Relativity of formality and informality in mathematics}
If we deal with some objects and try to interpret them as objects in some category, we would like to be able to deal with sets of the objects or sets of sets and so on. Simply, the objects should be freely arranged in sets with respect to some (meta) ZFC. These kinds of requirements are basic ingredients of what one may consider as a classical meta-environment for seeing the objects classically. So, being classical on the metalevel would mean being organized according to a set-like way. This would mean that the outer view of the category of the objects in question is regarded as classical provided it resembles the category SET. 

To be more specific let us formulate the following working definitions:
\begin{itemize}
\item One says that the metalevel is {\it organized classically} if it is formally modelled by SET.
\item {\it Formally modelled by} SET means the objects can be freely arranged into objects satisfying SET axioms. 
\item {\it Freely arranged in sets} can be understood in terms of set--based diagrams and its all finite limits or colimits which should exist, with the requirement that this outside SET structure agrees with the point structure of the objects while seen from the outside.   
\item We say a formal language is {\it classical at the level of symbols} if the symbols of the language are formally modelled by SET. The symbols of the language are interpreted in SET rather than in a general not two--valued and not classical topos. Thus, every symbol may have its name in some formal presentation with the use of classical sets similarly as the symbols are considered as separated and global in the preformal outside presentation. 
\item Similarly, we can say that a formal language is {\it intuitionistic at the level of symbols} provided the symbols of the language are internally interpreted in a topos. This is precisely the case where language or theory are interpreted in a topos \cite{moerdijkMcL1992}.
\item {\it Resemblance of some category with SET} is to be understood as possesing almost all or all categorical properties of SET by some other category. The specific list of the categorical properties should be specified. Isomorphism between SET and some other category is the example of the resemblance of categories.   
\end{itemize}
It is possible that a category resembles SET but does not respect the set--like structure of its objects. There are some toposes which have the categorical properties of SET, but the point--like structure of their objects is not preserved by morphisms of the toposes; the morphisms define elements and subobjects in the sense of the category but not in the sense of SET.     

Let us consider, as the example, some non classical topos with a natural number object (NNO). Objects of the topos can be organized twofold:
\begin{itemize}
\item classically, as the objects ''from the outside'' usually have classical set--theoretical structures, the one which agrees with classical meaning of element of a set and the description of them respects these structures.
\item nonclassically or internally, according to the internal, not SET--like and non classical, topos structure. The topos structure enforces the understanding of being an element (member of a set) or being subobject (subset) etc.. In general, any topos generates internal set theory \cite{moerdijkMcL1992};  
\end{itemize}
Even though the inherent nonclassicality derives from the not--SET and non classical structure of the topos exists, there is always the meta--level, such that the language of the theory of an elementary topos is the first order one. At this level one perfectly recognizes the symbols of the language as being arranged according to SET--like structures. The language is classical at the level of symbols. Moreover, at this level one can see the objects of the topos from the outside as having classical set-theoretical structures. 
   
The basic intuition which is of our concern here is that in any kind of formal presentation of a theory there always exists a sufficiently high level of the metatheory which would be organized classically and SET--like. That means that the formal language of a metatheory would be classical at the level of symbols.

Although the intuition behind this statement supports its obviousness, we attempt to argue against it in the following sections.    

Let $L$ be some formal language, of the first or the higher order, possibly with sorts $\mathbb{R}$ or $\mathbb{N}$ for real and natural numbers, and $T_M$ be a theory in the language which describes the manifold $M$. 

To have formally presented all the manifolds from the category $\cal{M}$ we have to associate to any manifold $M\in \cal{M}$ some formal language $L_M$ and the respective theory $T_M$ is formulated in this language.

If $M$ is considered as to be the universum of the model of $T_M$ we try to add nonlogical constant symbols to the language $L_M$ such that they exactly correspond to points of the manifold $M$. The intended interpretations of the constants in the model $M$ are just the points the constants correspond to. In this way we have what is refered to as a {\it simple extension} of the language. 

We say that a formal language $L$ {\it corresponds literally} to some open subset $U$ of some manifold $M$ if the language has in the set of its constant nonlogical symbols the ones corresponding uniquelly to the all points from $U$. Then we use a symbol $L_U$ for such a language. 

To see all the languages $L_U$ as sets of symbols we need some meta ZFC or PA to refer to, and $\mathbb{R}$ as a sort is understood classically with respect to this ZFC as well. Thus, every real from $\mathbb{R}$ may have a global name in the complex. 

It is assumed that the reference to such a ZFC or PA is always possible, though in an informal way, and that this is the absolute feature of a sufficiently rich metatheory.  

The family of sets of symbols $\{|L_U|\}$ determines the family of formal languages $\{L_U\}$ associated to the category $\cal{M}$ and which are simple extensions of the language $L$.
\begin{lemma}\label{lem2}
The family $\{|L_U|\}$ of sets of constant symbols corresponding to the formal languages $\{L_U\}$ is organized exactly as $Sh({\cal{M}})$ that is, the sets of symbols behave as sheaves with respect to morphisms of the category $\cal{M}$.
\end{lemma}
This is in fact expression of the well-known fact that the family of open subsets of a manifold form a Heyting algebra (etal\`e) rather than a Boolean algebra \cite{moerdijkMcL1992} and that the family of symbols of the languages should respect functorial morphisms in $\cal{M}$ when taking finer open covers. 

Now we know \cite{ReyesMoerdijk} that there are full and faithful embeddings of $\cal{M}$ in $\mathbb{L}$, in $\rm{SET}^{\mathbb{L}^{op}}$ and in $Sh(\mathbb{L})$. This means that 
\begin{lemma}\label{lemma2}
The formal language which would describe literally open domains of manifolds from $\cal{M}$ can be regarded as interpreted already in the topos $Sh(\mathbb{L})$ (it cannot be classical at the level of symbols). 
\end{lemma}
This is in fact reformulation of previous lemma and it states the fact that the families of sets of symbols are organized in the sheaves and $Sh(\mathbb{L})$ allows for the SET--like external organizing the symbols.
$\square$

Taking extensions of the theory still allowing for the literal descriptions of the manifolds, as having constant symbols in the language literally corresponding to the open domains, gives the following   
\begin{corollary}
In the case of the category of smooth manifolds and smooth maps between them any formal extension of the theory which would allow for literal descriptions of open domains of the objects of the category has to be intuitionistic. 
\end{corollary}
This is because the formal and literal description in question has to be intuitionistic at the level of the language symbols.

The peculiarity of this corollary is in that the category in question is the one whose objects are well-defined classical objects, namely smooth real finite dimensional manifolds, and they certainly should have a classical language suitable for a formal description of them. But while we attempt to keep the classical outside view of all the manifolds, the functional incompleteness of the category of manifolds become important and in the case of literal formal descriptions the language cannot be classical any more.  
This is the theory--metatheory complex which allows to see the language symbols as sets.

If we instead allow less formal connection between sets of language symbols and objects in the category (as
usually is the case) then the informality compensates nonclassical but formal character of the language symbols in question, and the symbols can be considered as classical sets. Moreover, the use of sorts $\mathbb{R}$ or $\mathbb{N}$, requires classical and set--like understanding of the sorts, if formally grasped, hence any element of the sorts should have its global name. Thus, extending the languages describing the manifolds toward literal ones is quite natural requirement while the sorts are in use. 
\begin{corollary}
Any higher order extension of the theory, describing literally open domains of the manifolds from $\cal{M}$, has to be performed internally in the topos of sheaves of the language symbols. 
\end{corollary}
This is just the reformulation of Lemma 2.
$\square$

If, instead, one considers the formal theory of manifolds which is not literal in the sense above, the extensions of such a theory can be classical and are not necessarily performed internally in the topos. In fact, literal formal presentation of the objects of the category and the language classical at the level of symbols cannot go together. 

If the objects of the category are considered formally and classically according to some meta ZFC then the languages describing them have to be well-defined in the sense that they have to have sets of symbols given classically. But in the case of topos--like modeling of the symbols of the languages we do not have the symbols as classical entities. Moreover, the topos can be described formally in the first order language classical at the level of symbols. Thus, the following holds: 

{\it The metalevel in the theory--metatheory complex has a formal intuitionistic description, extending the literal theory with some topos of sheaves of the symbols of the formal languages but this description is non-simultaneous with the formal descriptions of the objects of the category given by the languages.} 

Let us note that in the case when the metalevel is not a formal extension of the theory, the relation of the theory and the metatheory has to deal with some non formal ingredients. In this situation there is a possibility that the metalevel is presented formally as the formal topos and expressed in the language classical at the level of symbols. This presentation can be performed along with the formal presentations of the objects of the category. In this way we have some means allowing to distinguish between formal connection of the metalevel and the theory and a nonformal connection. Here is where informal ingredients still have to be presented in the complex.

Hence, we have the higher order extensions of the literal theory of manifolds from $\cal{M}$ constructed internally in the topos, or else classical SET--like environment where manifolds are arranged classically as language symbols do. 

In the first case, the metatheory is topos theory and, if formal, also this metatheory has its symbols organized classically. In the second case the metatheory is just ZFC or part of it. So, there is the classical ZFC meta-environment where classical manifolds are not within the reach of a literal and formal description. In the opposite case there is the intuitionistic meta-environment where manifolds can be reached formally and literally. 

One can say that when informal ingredients are allowed, the manifolds are seen classically but, when formally and literally, they have to be presented internally in the topos. In fact, we claim that both situations are not independent as purely objective, well-separated pictures of the same unaffected reality. As was conjectured in \cite{krol2004}, there are objects in the category $\cal{M}$ which require both of the pictures being entangled.

\section{\label{sec-12}Intuitionistic versus classical metaenvironment. Dynamics of the changes}
Let us, following MacLane and Moerdijk \cite{moerdijkMcL1992}, recall some main concepts connected with a language interpreted in a topos. 

Given any topos $\cal T$ we can specify the language of the topos which would allow us to built internally set--like constructions and it could express the true statements about the topos. The language is called the Mitchell--Benabou language. All constants and variables as well all formulas are of some type. The types are considered as objects of the topos $\cal T$. For each type $X$ there are variables of this type and they are represented by the identity morphism $1:X\rightarrow X$. 

A term $t$ of type $X$, possibly containing some free variables $y,z,w$ of types $Y,Z,W$, is represented by an arrow in $\cal T$ 
\begin{center}
$t:Y\times Z\times W \rightarrow X$
\end{center}
We do not present here detailed inductive definition of terms and their interpretations in a topos, since we do not use them explicitly. Again, all details can be found in \cite{moerdijkMcL1992}.

Terms $\phi , \psi , ...$ of type $\Omega_{\cal T}$ (subobjects classifier in the topos ${\cal T}$) are {\it formulas} of the language. We can connect formulas with various logical connectives and apply the quantifiers which yield formulas again of type $\Omega_{\cal T}$. This is possible, since the logical connectives are simply Heyting algebra operations defined on $\Omega_{\cal T}$, and can be composed with the morphisms ending on $\Omega $ as is in the case of formulas.

Similarly, the definition of quantifiers is possible internally so that while the quantifiers are applied to the formulas it results in the morphisms composition again of type $\Omega_{\cal T}$.  
Now, let us slightly modify the above $Sh(\mathbb{L})$ category such that the Grothendieck topology $J$ on it is going to be changed into the one which is also subcanonical (all representable presheaves from the Yoneda embedding are sheaves) and NNO is the one which contains infinite nonstandard naturals as well. This kind of modification is possible and relies only on the changes in the topology leaving the objects in $\mathbb{L}$ unaffected. The resulting category of sheaves in the changed topology on $\mathbb{L}$ is called Basel topos ($\cal B$) and it was described in details in Moerdijk and Reyes \cite{ReyesMoerdijk}.

Again, the sheaves in $\cal B$ can be considered as sheaves of language symbols, but with respect to new covering families deriving from the changed topology. The classes of sets of symbols which give descriptions of objects of $\mathbb{L}$ are organized as sheaves on the category $\mathbb{L}$. This results in the structure of Basel topos $\cal B$.

The change in the topology allows us to take as covers for the topology the ones which generate the Grothendieck topology as before, but with the addition that all projections along all nontrivial loci generate the covers as well \cite{ReyesMoerdijk}.    

Now, we attempt to be more explicit in demonstrating the role played by interpreted language in the topos while the special kind of classical objects is considered. The basic observation is that the object of natural numbers in $\cal B$ contains infinite ones. The classical counterpart of this is the existence of nonstandard models of PA (in the sense of Robinson) which also contain the infinite big natural numbers. Furthermore, the object of smooth real numbers in $\cal B$ contains (intuitionistically) indempotent reals $d^2=0$ and infinite big and infinite small ones. Classically indempotents reals do not exist in any model of real numbers but again, there are Robinsonnian models of real numbers which are classical and which contain nonstandard big and small numbers. Thus, one can consider the nonstandard models of real numbers as a classical counterpart of intuitionistic object of smooth real numbers. One can say that the object of {\it smooth} NN and {\it smooth} real numbers from $\cal B$ are projected on classical domains (or projected classically), namely on the ${\bf ^\ast N}$ and ${\bf ^\star R}$ respectively.

Let us introduce some working definitions. 
\begin{itemize}
\item {\it An object exists classically} provided its existence is verified by some classical theories. 
\item {\it Classical theories} are those which are based on classical logic but they are not necessarily axiomatized or even purely formal. They should be considered as the complex {\it theory -- metatheory} which gives mathematical results (theorems). 
\item A formal language, $L({\rm SET},{\cal B})$, is called to be {\it dynamically changing from ${\rm SET}$ to ${\cal B}$} if among its symbols are those from ${\rm SET}$ and others internally interpreted in ${\cal B}$.  
\item We say that an object which exists classically {\it is described generically by a dynamical language} $L({\rm SET},{\cal B})$ if the object has to refer both to the numbers from $\mathbb{N}$ ($\mathbb{R}$) and from ${\bf ^\ast N}$ (${\bf ^\ast R}$) by the symbols of the language, but ${\bf ^\ast N}$, ${\bf ^\ast R}$ are considered as projected classically from ${\cal B}$.
\end{itemize}
\begin{lemma}\label{lem33}
If the meta-language is considered as changing dynamically from the constant SET-like into the one interpreted internally in Basel topos, and if there exists a classical object described generically by the varying language, then there exist two models of PA, one standard and one nonstandard given by the ultrafilter construction, which are not formally distinguished from one another. This means that there does not exist any higher order property of standard natural numbers, which would not be valid in the nonstandard model.
\end{lemma}
This is possible since the higher order properties are defined with respect to SET and compared with those defined with respect to the internal language of ${\cal B}$. If in SET, standard model, $\mathbb{N}$, of PA appears. If internally in $\cal B$, then classical projection of $smooth$-$N$ gives ${\bf ^\ast N}$. ${\bf ^\ast N}$ is given classically by the ultrafilter construction. 
 
Based on Lemma \ref{lem33}, we have

{\it Under assumptions of Lemma \ref{lem33}, the standard model of real field and some nonstandard one given by the ultrafilter construction are indistinguishable. This means that there does not exist any higher order property of standard real numbers, which would not be valid in the nonstandard model.  
This indistinguishability has a formal meaning for some smooth manifolds from $\cal{M}$, namely for exotic $R^4$'s.}

Similarly, one cannot distinguish the standard real numbers and nonstandard ones, by any higher order property, provided the quantification is performed internally i.e. over internal subsets of ${\bf ^\ast R}$ \cite{kock1974}.

Formal means needed to distinguish the models are already internally interpreted in $\cal B$. On the other hand, informal ingredients have to be always present in the complex theory--metatheory, and these sometimes cause lack of formal control over the meta-environment. The possible meaning of the indistinguishability $\mathbb{R}^4$ and ${\bf ^\star R^4}$ for some (classical) manifolds is conjectured \cite{krol2004,krol2004b}. The smooth structures of the manifolds can survive the shift. Exotic $R^4$'s are certainly classically existing entities, since the proof of the existence of these was obtained classically.

Equivalently, in terms of invariant mathematics, i.e. valid in all toposes with NNO, one can say that formal differences between local presentations with respect to the specific frames-toposes are neglected. Only properties valid in all toposes with NNO are expressible invariantly. In particular, in ${\cal B}$ NNO is $s$-$N$ and in SET this is standard $\mathbb{N}$ and the invariance requires that no difference between them is expressed. This indistinguishability, however, can have in turn, formal back-reaction. Thus, the shift from SET to $\cal B$ has a price in mathematics; it is the emergence of exotic $R^4$'s. And this is like the emergence of gravitational effects in the covariant, tensorial, language of GR. 

One can also say that there are some smooth manifolds requiring both classical and intuitionistic meta-levels if formal.

The indistinguishability of the models of PA or real numbers is refered to as $main$ $hypothesis$ in what follows.

To be more specific, let the types of the language symbols be sheaves in $Sh({\mathbb{L}})$ or ${\cal B}$. It means that the variables and all terms are interpreted by suitable morphisms in the topos ${\cal B}$. All formulas are as usual of type $\Omega _{\cal B}$ (classifying object of ${\cal B}$).

The entire interpretation is forced by  
\begin{itemize}
\item Taking constant symbols as being of the type of some sheaves in the topos. 
\item Respecting the SET--like, classical grouping of the symbols of the language i.e. respecting the inverse limits etc. as in the categorical presentation.
\end{itemize}
Thus, the variables are also of the ${\cal B}$--sheaves types. Subsequently, one has to interpret the whole language in the topos. Let us note that a topos can serve as a model for any higher order language. 

In this way we can have all set--like classes as $\{x:\phi (x)\}$ already in the topos, and the validity of formal statements in the topos can be translated into the outside one simply by avoiding the excluded midle law and the axiom of choice in the argumentation performed in the internal language. In fact, we have a language of the invariant mathematics. Being equipped with such an internal language in ${\cal B}$ we have to use it in the context of smooth manifolds which are formally and literally approached. 
 
The special objects can exist which are able {\it to survive as classical objects} the shift in the descriptions given by constant classical language and the language already interpreted in the topos as above. In the case of smooth manifolds this would result in the situation where both descriptions should be connected by some diffeomorphisms if the object survives as the classical object. Since the diffeomorphism cannot be standard, this may serve as an approach toward the understanding of the so called {\it exotic diffeomorphisms} in the case of exotic smooth $R^4$'s as in \cite{krol2004b}.  
\newpage 
\acknowledgments{I would like to thank professor Robert Gompf for the correspondence and explaining to me many points regarding exotic $R^4$'s. I owe special thanks to the organizers of the IPM String School and Workshop for their hospitality and giving to me the chance to participate in the School. }

\end{document}